\documentstyle[12pt,cite]{article}
              
\title{Variance of the Quantum Coordinates \\ of an Event.}
\author{M. Toller   \\ 
Department of Physics of the University and I.N.F.N.  \\
I-38050 Trento, Italy}
 
\begin{document} 
\maketitle                             
                 
\begin{abstract}
We study the variances of the coordinates of an event considered as quantum observables in a Poincar\'e covariant theory.  The starting point is their description in terms of a covariant positive-operator-valued measure on the Minkowski space-time.  Besides the usual uncertainty relations, we find stronger inequalities involving the mass and the centre-of-mass angular momentum of the object that defines the event.  We suggest that these inequalities may help to clarify some of the arguments which have been given in favour of a gravitational quantum limit to the accuracy of time and space measurements.

\bigskip  \bigskip

\noindent PACS: \quad 03.65.-w - quantum theory; \quad 03.30.+p - special relativity; 

04.60.-m - quantum gravity.

\end{abstract}
\newpage

\section{Introduction.}  

The word ``event'' is often used to indicate a point of the space-time manifold. From the operational point of view this point has to be defined in terms of some properties of a material object and the word ``event'' assumes a meaning which is not purely geometric but is more similar to the meaning it has in the usual language, namely ``something that happens'' in the physical world. A typical event is the collision of two particles. Since the collision of two small particles has a small probability, one may consider, more precisely, in the rest system of the centre-of-mass (CoM), the point which coincides with the CoM when the distance between the two particles reaches its minimum value. This definition of an event makes sense in a classical (non-quantum) theory in the absence of strong gravitational fields, which would complicate the space-time geometry. In the example given above we have considered a ``baricentric event'', which lies on the world line of the CoM of the object that defines it. There are also non-baricentric events: for instance a collision of the first two particles in a system of three particles.  

If we take into account the quantum properties of the object that defines an event, we have to face some difficult problems. The space-time coordinates $x^{\alpha}$ $(\alpha = 0, 1, 2, 3)$ of the event with respect to a classical reference frame have to be considered as quantum observables. The corresponding operators $X^{\alpha}$ have been defined and discussed in refs.\ \cite{JR, JR2, JR3, JR4} in the framework of a conformally covariant quantum theory. A detailed treatment of events in the framework of a Poincar\'e covariant quantum theory is given in ref.\ \cite{Toller} and the present article is a continuation of this research. Since the Hermitian operators $X^{\alpha}$ cannot be self-adjoint \cite{Wightman, Pauli}, a complete description of the coordinate observables has to  be given in terms of a positive-operator-valued measure (POVM) \cite{Davies, Holevo, Werner, BLM, BGL, BGL2, Giannitrapani} defined on the Minkowski space-time  $\cal M$ and covariant with respect to the Poincar\'e group. An explicit formula for the most general Poincar\'e covariant POVM is given in ref.\ \cite{Toller}; it permits the calculation of the operators $X^{\alpha}$.  

Note that different measurement procedures are described by different POVMs and there is in principle no reason to concentrate the attention on one of them. On the other hand, one cannot assume that every covariant POVM describes a (possibly idealized) measurement procedure. The POVM formalism is based on very general principles of relativity and quantum theory, but other requirements may be relevant, for instance superselection rules or discrete symmetries as time-reversal. As a consequence, some care is needed in the physical interpretation of our results.

If the CoM angular momentum does not vanish, the uncertainty relations do not permit the definition an exactly baricentric event in a quantum theory, but one can define an important class of POVMs, called quasi-baricentric, which describe events which happen as near as possible to the world-line of the CoM.  From a POVM of this kind, assuming its covariance under dilatations, one obtains exactly the operators $X^{\alpha}$ defined and justified with good arguments in refs.\ \cite{JR, JR2, JR3, JR4}. In Section 3 we characterize the quasi-baricentric POVMs by means of another physically relevant condition.

If the state of the system is described by a vector $\psi \in {\cal H}$, the POVM $\tau$ defines the probability that the event is detected in a set $I \in {\cal M}$. Actually, one can write this probability in the form   
\begin{equation}  
(\psi, \tau(I) \psi) = \int_I \rho(\psi, x) \, d^4 x, 
\end{equation} 
where the density $\rho(\psi, x)$ is an integrable function. We have also shown in \cite{Toller} that the operators $\tau(I)$ have to vanish on the states with a singular four-momentum spectrum, namely the vacuum and the one-particle states. Therefore we consider a Hilbert space ${\cal H}$ that contains only states with a continuous mass spectrum, for instance scattering states with two or more incoming or outgoing particles. In the following we assume that $\tau$  is normalized, namely that
\begin{equation}  \label{Normalization}
\tau({\cal M}) = \int_{\cal M} \rho(\psi, x) \, d^4 x = 1, \qquad \psi \in {\cal H}. 
\end{equation} 
This means that the event has to be found somewhere in space-time. 

The average value of a coordinate is given by
\begin{equation} \label{Average1}  
\langle x^{\alpha} \rangle = \int x^{\alpha} \rho(\psi, x) \, d^4 x = (\psi, X^{\alpha} \psi). 
\end{equation}
This equation defines the Hermitian coordinate operators $X^{\alpha}$, which in general do not commute. Note, however, that the averages
\begin{equation} \label{Average2}   
\langle (x^{\alpha})^2 \rangle = \int (x^{\alpha})^2  \rho(\psi, x) \, d^4 x
\end{equation}
in general cannot be written in the form     
\begin{equation}  \label{Wrong}
\langle (x^{\alpha})^2 \rangle =  (\psi, (X^{\alpha})^2 \psi) = \|X^{\alpha} \psi\|^2
\end{equation} 
and that the operators $X^{\alpha}$ do not determine the POVM $\tau$ univocally. There are translation covariant POVMs that satisfy eq.\ (\ref{Wrong}), but this is not possible if the Poincar\'e covariance is required. A general discussion of this problem (in a different context) can be foumd in \cite{Werner}. It is important to remember the the integrals (\ref{Average1}) and (\ref{Average2}) do not converge for all the choices of the vector $\psi$ and that the unbounded operators $X^{\alpha}$, as many other quantum observables, are not defined on the whole Hilbert space $\cal H$.

The purpose of the present article is to give a treatment of the quantities $\langle (x^{\alpha})^2 \rangle$ by means of the formalism developed in ref.\ \cite{Toller}. We can always choose a reference frame in which $\langle x^{\alpha} \rangle = 0$ and in this frame  $\langle (x^{\alpha})^2 \rangle$ is just the variance $(\Delta x^{\alpha})^2$. It is also possible to choose the reference frame in such a way that the positive definite matrix $\langle x^{\alpha} x^{\beta} \rangle$ is diagonal.  In Section 2 we show that the indeterminacy relations 
\begin{equation}  \label{Indet} 
\Delta x^{\alpha} \Delta k^{\alpha} \geq \frac 1 2, \qquad c= \hbar = 1, 
\end{equation}  
where $k$ is the four-momentum of the system that defines the event, are valid in general, even if the proof is more complicated than the usual one.

Then we find other inequalities which involve the square of the CoM angular momentum which we indicate by ${\bf J}^2 =j(j+1)$. For instance, we find 
\begin{equation} \label{Ineq} 
\sum_{r=1}^3 (\Delta x^r)^2 \geq \sum_{r=1}^3 (2 \Delta k^r)^{-2} + \langle \theta_j (j+1) \mu^{-2} \rangle,
\end{equation}   
where
\begin{equation} 
\mu = (k_{\alpha} k^{\alpha})^{1/2}
\end{equation}   
and  
\begin{equation} 
\theta_0 = 0, \qquad \theta_j = 1 \quad {\rm for} \quad j > 0.
\end{equation}  
The inequality (\ref{Ineq}), however, is valid only in the absence of interference between terms with different values of $j$. For POVMs of the most general kind, this happens if $\psi$ is an eigenvector of ${\bf J}^2$. Alternativey one can consider an arbitrary vector $\psi$ and require that the operators $\tau(I)$ are diagonal in the index $j$, namely they commute with ${\bf J}^2$. The quasi-baricentric POVMs and the corresponding operators $X^{\alpha}$ have this property. Note that the inequality (\ref{Ineq}) does not follow in the usual way from the commutation relations \cite{JR3} between the operators $X^{\alpha}$. The inequalities obtained in this way contain the averages of the components of the vector ${\bf J}$, which can vanish even for eigenstates of ${\bf J}^2$ with a large eigenvalue $j(j+1)$.

In Section 3 we consider the quasi-baricentric POVMs with more detail. We find an example in which $\Delta x^0$ and $\Delta x^3$ can be arbitrarily small, even if $\langle \theta_j (j+1) \mu^{-2} \rangle$ takes an arbitrary fixed value.  In Section 4 we suggest that, for two-particle states, the quantity $j \mu^{-2}$ can be used to control the appearance of strong gravitational fields, which cannot be treated within the range of validity of a Poincar\'e covariant quantum theory. Then we indicate how our inequalities can be used to discuss in a more formal way some aspects of the quantum gravitational limitations to the accuracy of the measurements of time and position. 

We hope that our results will contribute to show that the POVM formalism is not just a trick to avoid some  mathematical inconsistencies, but it can reveal interesting and perhaps unexpected physical effects.

\section{Calculation of the variances.} 
                                        
A vector $\psi \in {\cal H}$ is described by a wave function of the kind  $\psi_{\sigma j m}(k)$, where $k$ is the four-momentum, $j, m$ describe the angular momentum in the CoM frame and the index $\sigma$ summarizes all the other quantum numbers. For instance, in a two-particle state $\sigma$ describes the CoM helicities \cite{JW}. Since $\psi$ has a continuous four-momentum spectrum, its norm can be written in the form
\begin{equation}  
\|\psi\|^2 = \int_V \sum_{\sigma j m} |\psi_{\sigma j m}(k)|^2  \, d^4 k,  
\end{equation}
where $V$ is the open future cone. Note that $j$ and the mass $\mu$ label the equivalence classes of irreducible unitary representations of the Poincar\'e group, which act on the wave functions $\psi_{\sigma j m}(k)$ in the way described by Wigner \cite{Wigner}. 

We have shown in \cite{Toller} that the probability density is given by
\begin{equation} 
\rho(\psi, x) =  \int_{\Gamma} \sum_{ln} |\Psi_{\gamma l n}(x)|^2 \, d \omega(\gamma),
\end{equation}
where                  
\begin{equation} 
\Psi_{\gamma ln}(x) = (2 \pi)^{-2}  \int_V \exp(-i x_{\alpha} k^{\alpha}) \sum_{jm} D^{M c}_{lnjm}(a_k) \phi_{\gamma jm}(k) \, d^4 k.
\end{equation} 
We have introduced the new function
\begin{equation} 
\phi_{\gamma jm}(k) = \sum_{\sigma} F_{\gamma \sigma}^j(\mu) \psi_{\sigma jm}(k).
\end{equation}
The quantities $F^j_{\gamma \sigma}(\mu)$ are complex functions of $\mu$ which characterize the particular POVM. The quantities  $D^{M c}_{jmj'm'}(a)$ are the matrix elements of the irreducible unitary representations of $SL(2, C)$ \cite{Naimark, GGV, Ruhl} and the possible values of the indices  are
\begin{displaymath}
M = 0, \pm \frac 1 2, \pm 1,\ldots, \qquad  c^2 < 1,
\end{displaymath}
\begin{equation} 
j = |M|, |M| + 1,\ldots, \qquad  m = -j, -j + 1,\ldots, j.
\end{equation}
If $M \neq 0$, $c$ must be imaginary. The representations $D^{Mc}$ and $D^{-M, -c}$ are unitarily equivalent. According to our conventions, $D^{01}$ is the trivial one-dimensional representation with only one matrix element given by
\begin{equation} 
D^{01}_{0000}(a) = 1.
\end{equation}  
The variable $\gamma \in \Gamma$, stands for a discrete index $\nu$ and the parameters $M$, $c$ which label the irreducible unitary representations of $SL(2, C)$; $\omega$ is a positive measure on the space $\Gamma$.  For the Wigner boost \cite{Wigner} we use the choice  
\begin{equation}   
a_k = (2 \mu (\mu + k^0))^{- 1/2} (\mu + k^0 + k^s \sigma^s) \in SL(2, C),
\end{equation}
where $\sigma^s$ are the Pauli matrices. Here and in the following the indices $\alpha, \beta$ take the values $0, 1, 2, 3$ and the indices $r, s, t, u, v$ take the values $1, 2, 3$. The summation convention is applied to both these kinds of indices.  The normalization condition (\ref{Normalization}) gives rise to the constraint
\begin{equation}  \label{Normalization2}
\int_{\Gamma} \overline{F_{\gamma \sigma}^j(\mu)} F_{\gamma \sigma'}^j(\mu) \, d\omega(\gamma) = \delta_{\sigma \sigma'}
\end{equation}
and to the equation
\begin{equation} 
\int_{\Gamma} \int_V \sum_{jm} |\phi_{\gamma jm}(k)|^2 \, d^4 k \, d \omega(\gamma) = \|\psi\|^2.
\end{equation}   

In order to compute the averages (\ref{Average1}) and (\ref{Average2}), we note that we have, in the sense of distribution theory, 
\begin{displaymath} 
x_{\alpha} \Psi_{\gamma ln}(x) =
\end{displaymath}
\begin{equation} 
= -i (2 \pi)^{-2} \int_V \exp(-i x_{\alpha} k^{\alpha}) \sum_{jm} \frac{\partial}{\partial k^{\alpha}} \left( D^{M c}_{lnjm}(a_k) \phi_{\gamma jm}(k) \right) \, d^4 k. 
\end{equation}
In this way we obtain, if $\psi$ is chosen in such a way that the integral is meaningful and convergent, 
\begin{equation}  \label{Variance}
\langle (x^{\alpha})^2 \rangle = \int_{\Gamma} \int_V \sum_{jm} |[Y^{\alpha} \phi]_{\gamma jm}(k)|^2 \, d^4 k \, d \omega(\gamma),
\end{equation}
where the operators $Y^{\alpha}$ are defined by
\begin{equation} 
[Y_{\alpha} \phi]_{\gamma jm}(k) = \sum_{j'm'} S^{M c}_{\alpha jmj'm'}(k) \phi_{\gamma j'm'}(k) - i \frac{\partial}{\partial k^{\alpha}} \phi_{\gamma jm}(k)
\end{equation} 
and we have introduced the Hermitian matrices
\begin{equation} 
S^{M c}_{\alpha jmj'm'}(k) = -i \sum_{ln} D^{M c}_{jmln}(a_k^{-1}) \frac{\partial}{\partial k^{\alpha}} D^{M c}_{lnj'm'}(a_k).
\end{equation} 

These matrices, as we have shown in ref.\ \cite{Toller}, are given by
\begin{equation} 
S^{M c}_{0 jmj'm'}(k) = \frac{1}{\mu^2} k^r N^{rMc}_{jmj'm'},  
\end{equation}   
\begin{displaymath} 
S^{M c}_{r jmj'm'}(k) = - \frac{1}{\mu} N^{rMc}_{jmj'm'} -
\end{displaymath}
\begin{equation}
- \frac{1}{\mu^2 (\mu + k^0)} k^r k^s N^{sMc}_{jmj'm'} + \frac{1}{\mu (\mu + k^0)} \delta_{jj'} \epsilon^{rst} k^s M^{tj}_{mm'}.
\end{equation}
We have indicated by  $M^{r j}_{mm'}$ the usual Hermitean angular momentum matrices and  $N^{rMc}_{jmj'm'}$ are the Hermitian generators of the boosts in the representation $D^{M c}$ of $SL(2, C)$. They  vanish unless $j - j' = 0, \pm 1$ and can be found (with different notations) in ref.\ \cite{Naimark}. For $M = j =0$ and $c = 1$, all these quantities vanish.  Note that  
\begin{equation} 
k^{\alpha} S^{M c}_{\alpha jmj'm'}(k) = 0.
\end{equation}   

The Hermitian operators $Y^{\alpha}$ operate in a Hilbert space larger than the physical Hilbert space $\cal H$ and should not be confused with the coordinate operators $X^{\alpha}$. Nevertheless, if we indicate by $K^{\beta}$ the multiplication by $k^{\beta}$, they satisfy the commutation relations 
\begin{equation} 
[Y^{\alpha}, K^{\beta}] = -i g^{\alpha \beta}.
\end{equation}                                
Then, by means of the usual procedure based on the Schwarz inequality and working in a reference frame in which $\langle x^{\alpha} \rangle = 0$, we obtain the inequalities (\ref{Indet}), where
\begin{displaymath} 
(\Delta k^{\alpha})^2 = \int_{\Gamma} \int_V \sum_{jm} (k^{\alpha} - \langle k^{\alpha} \rangle)^2 |\phi_{\gamma jm}(k)|^2 \, d^4 k \, d \omega(\gamma) =
\end{displaymath}  
\begin{equation}  
= \int_V \sum_{\sigma jm} (k^{\alpha} - \langle k^{\alpha} \rangle)^2 |\psi_{\sigma jm}(k)|^2 \, d^4 k 
\end{equation}
is the variance of a component of the physical four-momentum.

In order to find more restrictive inequalities, we consider a wave function $\psi_{\sigma jm}(k)$ which does not vanish only for a given value of the index $j$. Then  eq.\ (\ref{Variance}) can be written more explicitly in the form 
\begin{displaymath} 
\langle (x^{\alpha})^2 \rangle = \int_{\Gamma} \int_V \sum_{m} |[Z^{\alpha} \phi]_{\gamma jm}(k)|^2 \, d^4 k \, d \omega(\gamma) +
\end{displaymath} 
\begin{displaymath} 
+ \int_{\Gamma} \int_V \sum_{m} |\sum_{m'} S^{M c}_{\alpha, j+1,m,j,m'}(k) \phi_{\gamma jm'}(k)|^2 \, d^4 k \, d \omega(\gamma) +
\end{displaymath}
\begin{equation}  \label{Variance2}
 + \int_{\Gamma} \int_V \sum_{m} |\sum_{m'} S^{M c}_{\alpha, j-1, m, j, m'}(k) \phi_{\gamma jm'}(k)|^2 \, d^4 k \, d \omega(\gamma).
\end{equation} 
where
\begin{equation} \label{Zeta}
[Z_{\alpha} \phi]_{\gamma jm}(k) = \sum_{m'} S^{M c}_{\alpha jmjm'}(k) \phi_{\gamma jm'}(k) - i \frac{\partial}{\partial k^{\alpha}} \phi_{\gamma jm}(k),
\end{equation}                                           
namely $Z_{\alpha}$ is the part of $Y_{\alpha}$ which is diagonal with respect to the index $j$.
By means of the procedure used above, one can show that the first term in the right hand side has the lower boud $(2 \Delta k^{\alpha})^{-2}$ and the other two terms, which do not contain derivatives, improve this lower bound. 

From the results given in ref.\ \cite{Naimark} we obtain, in agreement with the Wigner-Eckart theorem, the following useful formulas
\begin{equation}
N^{rMc}_{jmjm'} = \frac{-i M c}{j (j + 1)} M^{r j}_{mm'},
\end{equation}
\begin{displaymath} 
\sum_{m''} N^{rMc}_{j,m,j+1,m''} N^{sMc}_{j+1,m'',j,m'} =
\end{displaymath}
\begin{equation}
= Q^{Mc}_{j+1} \left( (j+1)^2 \delta_{rs} \delta_{mm'} - \sum_{m''} M^{rj}_{mm''} M^{sj}_{m''m'} - i (j + 1) \epsilon^{rst} M^{tj}_{mm'} \right),
\end{equation}
\begin{displaymath} 
\sum_{m''} N^{rMc}_{j,m,j-1,m''} N^{sMc}_{j-1,m'',j,m'} =
\end{displaymath}
\begin{equation}
= Q^{Mc}_j \left( j^2 \delta_{rs} \delta_{mm'} - \sum_{m''} M^{rj}_{mm''} M^{sj}_{m''m'} + i j \epsilon^{rst} M^{tj}_{mm'} \right).
\end{equation}   
where
\begin{equation}
Q^{Mc}_j = \frac{(j^2 - M^2)(j^2 - c^2)}{j^2 (2j + 1)(2j - 1)}, \qquad j \geq 1.
\end{equation} 

From these formulas we obtain: 
\begin{equation}
\sum_{m''} S^{Mc}_{0,j,m,j+1,m''} S^{Mc}_{0,j+1,m'',j,m'} = \mu^{-4} Q^{Mc}_{j+1} A^j_{mm'}(k),
\end{equation}
\begin{displaymath} 
\sum_{m''} S^{Mc}_{r,j,m,j+1,m''} S^{Mc}_{r,j+1,m'',j,m'} =
\end{displaymath} 
\begin{equation}   \label{SS}
= \mu^{-2} Q^{Mc}_{j+1} (j+1) (2j + 3) \delta_{mm'}  + \mu^{-4} Q^{Mc}_{j+1} A^j_{mm'}(k),
\end{equation}
where     
\begin{equation}
A^j_{mm'}(k) = k^r k^r (j+1)^2 \delta_{mm'} - k^r k^s \sum_{m''} M^{rj}_{mm''} M^{sj}_{m''m'} 
\end{equation}  
is a positive definite matrix.

Then, in a reference frame in which $\langle x^{\alpha} \rangle = 0$,  by means of the inequality
\begin{equation}  \label{QMin}
Q^{Mc}_{j+1} \geq \theta_j (2j + 3)^{-1}, \qquad |M| \leq j,
\end{equation} 
we obtain the inequalitiy (\ref{Ineq}). We have to remember that it has been proven under the assumption that $\psi$ is an eigenvector of ${\bf J}^2$. Its general validity can be excluded: we have just to consider a non-baricentric POVM which describes a collision of the first two particles in a system of three particles. Then, if the third particle is sufficiently distant, the total centre-of-mass angular momentum $j$ can be made as large as we want without increasing the CoM energy $\mu$ and without affecting the variance of the coordinates of the considered event.

\section{Quasi-baricentric events.}  

In order to minimize the second term in eq.\ (\ref{Variance2}), we have to require that the equality sign holds in eq.\ (\ref{QMin}), namely that
\begin{displaymath} 
F^j_{\nu M c \sigma}(\mu) \neq 0 \qquad {\rm only\,\, if} 
\end{displaymath}   
\begin{equation}  \label{Baricentric}
M = j, \qquad c= 1 \quad {\rm for} \quad j = 0 \qquad {\rm and} \quad c = 0 \quad {\rm for} \quad j > 0. 
\end{equation}
Under the same conditions, the third term in eq.\ (\ref{Variance2}) vanishes. This is the definition of quasi-baricentric POVM introduced in ref.\ \cite{Toller} with a different motivation. We may say that the quasi-baricentric POVMs minimize the variances of the coordinates when $\psi$ is an eigenvector of ${\bf J}^2$.

If the condition (\ref{Baricentric}) is valid, the index $j$ is uniquely fixed by the index $\gamma = \{\nu, M, c \}$ and no interference term can appear. Then we expect that the inequality (\ref{Ineq}) derived in the preceding Section is valid for a quasi-baricentric measure without any limitation of the wave function $\psi$. In order to to study this inequality with more detail, we rewrite the formulas given in the preceding Section  in a simpler form, in particular we replace the integration over the variable $\gamma = \{\nu, M, c\}$ by a sum over the indices $\nu$ and $M = j$. In this way we obtain
\begin{equation} 
\rho(\psi, x) =  \sum_{\nu jln} |\Psi_{\nu jln}(x)|^2,
\end{equation}                 
\begin{equation} 
\Psi_{\nu jln}(x) = (2 \pi)^{-2}  \int_V \exp(-i x_{\alpha} k^{\alpha}) \sum_m D^{j c}_{lnjm}(a_k) \phi_{\nu jm}(k) \, d^4 k.
\end{equation} 
\begin{equation} 
\phi_{\nu jm}(k) = \sum_{\sigma} F_{\nu \sigma}^j(\mu) \psi_{\sigma jm}(k),
\end{equation}
where the quantities
\begin{equation} 
F_{\nu \sigma}^j(\mu) = F_{\nu j c \sigma}^j(\mu)
\end{equation} 
have the normalization property
\begin{equation}  \label{Normalization3}
\sum_{\nu} \overline{F_{\nu \sigma}^j(\mu)} F_{\nu \sigma'}^j(\mu) = \delta_{\sigma \sigma'}.
\end{equation}                                   
In all these formulas, the parameter $c$ takes the values given in eq.\ (\ref{Baricentric}).

Eq.\ (\ref{Variance}) can be written in the form
\begin{equation}  \label{Variance3}
\langle (x^{\alpha})^2 \rangle = \|Z_{\alpha} \phi\|^2 + \int_V \sum_{\nu jm} | \sum_{m'} S^{j c}_{\alpha,j+1,m,j,m'}(k) \phi_{\nu jm'}(k) |^2 \, d^4 k,
\end{equation} 
where 
\begin{equation} 
\|Z_{\alpha} \phi\|^2 = \int_V \sum_{\nu jm} |[Z_{\alpha} \phi]_{\nu jm}(k)|^2 \, d^4 k.
\end{equation} 
For a quasi-baricentric measure the operators $Z_{\alpha}$ defined by eq.\ (\ref{Zeta}) take the form 
\begin{equation} 
Z_0 = - i \frac{\partial}{\partial k^0},
\end{equation} 
\begin{equation} 
Z_r = \mu^{-1} (\mu + k^0)^{-1}\epsilon^{rst} k^s M^{tj} - i \frac{\partial}{\partial k^r}.
\end{equation}
where $M^{tj}$ are the angular momentum matrices which act on the index $m$ of $\phi_{\nu jm}(k)$.
These operators satisfy the commutation relations
\begin{equation} 
[Z_{\alpha}, K^{\beta}] = -i \delta_{\alpha}^{ \beta},
\end{equation}  
\begin{equation} 
[Z_0, Z_r] = i \mu^{-3} \epsilon^{rst} k^s M^{tj},
\end{equation}  
\begin{equation}  \label{Comm}
[Z_r, Z_s] = i \mu^{-3} \left(k^0 \epsilon^{rsu} -(\mu + k^0)^{-1} \epsilon^{rsv} k^v k^u \right) M^{uj}.
\end{equation}
These relations permit, in the usual way, to obtain inequalities for the quantities $\|Z_{\alpha} \phi\|$. In particular, using also eq.\ (\ref{SS}), one can derive the inequality (\ref{Ineq}).  

It is interesting to consider the case in which
\begin{equation} \label{Particular} 
F_{\nu \sigma}^j(\mu) = \delta_{\nu \sigma}, \qquad  \phi_{\nu jm}(k) = \psi_{\nu jm}(k).  
\end{equation}  
Then the operators $Z^{\alpha}$ operate in the physical Hilbert space $\cal H$ and they are just the coordinate operators $X^{\alpha}$. From eq.\ (\ref{Variance3}) we see that eq. (\ref{Wrong}) cannot be true, but there are additional contributions to the variance.

In order to obtain a better understanding of eq.\ (\ref{Variance3}), we examine a simple model. We consider a quasi-baricentric POVM which satisfies eq.\ (\ref{Particular}), we assume that the index $\sigma = \nu$ can take only one value and we drop it in the following formulas. Then we consider a particular sequence of vectors $\psi^{(j)}$ given by
\begin{equation} 
\psi^{(j)}_{j'm}(k) =  \delta_{j'j} \delta_{mj} f(q) j^{-3/8},
\end{equation} 
where
\begin{equation} 
q^0 = j^{-1/2} k^0, \qquad  q^1 = k^1, \qquad q^2 = k^2, \qquad q^3 = j^{-1/4} k^3, 
\end{equation} 
and we compute the limit of the variances for $j \to \infty$. In this limit the CoM angular momentum becomes very large and is directed along the $x^3$ axis, the variances of $k^0$ and $k^3$ tend to infinity in a different way and the components $k^r (k^0)^{-1}$ of the CoM velocity become very small.  By means of a change of variables, we see that the normalization condition of the vectors $\psi^{(j)}$ is given by
\begin{equation} 
\|\psi^{(j)}\|^2 =  \int |f(q)|^2 \, d^4 q = 1
\end{equation}
and that we have the finite limit
\begin{equation} 
\lim_{j \to \infty} \langle (j + 1) \mu^{-2} \rangle = \int |f(q)|^2 (q^0)^{-2} \, d^4 q = A.
\end{equation}                                

By means of the formulas given in the preceding Section and of the known form of the angular momentum matrices matrices $M^{rj}$, we obtain after some calculations
\begin{equation} 
\lim_{j \to \infty} \langle (x^0)^2 \rangle = \lim_{j \to \infty} \langle (x^3)^2 \rangle = 0,
\end{equation} 
\begin{equation}
\lim_{j \to \infty} \langle (x^1)^2 \rangle = \int |\frac 1 2 q^2 (q^0)^{-2} f(q) - i \frac{\partial}{\partial q^1} f(q) |^2  \, d^4 q + \frac 1 2 A,
\end{equation}
\begin{equation}
\lim_{j \to \infty} \langle (x^2)^2 \rangle = \int |-\frac 1 2 q^1 (q^0)^{-2} f(q) - i \frac{\partial}{\partial q^2} f(q) |^2 \, d^4 q + \frac 1 2 A.
\end{equation}  
We see that if we fix the quantity  $\langle (j + 1) \mu^{-2} \rangle$, the quantities $\Delta x^0$ and $\Delta x^3$ can be arbitrarily small.  The integrals in the last two formulas cannot be neglected with respect to $A$, as it follows from the commutation relations (\ref{Comm}). We can show that they can be made of the order of A by computing them for a special choice of $f(q)$, namely
\begin{equation}
f(q) = (2\pi)^{-1/2} (q^0)^{-1} \exp\left(-(2 q^0)^{-2} ((q^1)^2 + (q^2)^2)\right) g(q^0, q^3).
\end{equation}                                                                        
The result is
\begin{equation}
\lim_{j \to \infty} \langle (x^1)^2 \rangle = \lim_{j \to \infty} \langle (x^2)^2 \rangle = A.
\end{equation}

\section{Peripheral collisions.} 

Many arguments indicate that the quantum gravitational effects give rise to some limitations to the precision of time and position measurements (for a general review, see ref.\ \cite{Garay}).  These limitations are described by some inequalities involving the quantities $\Delta x^{\alpha}$ and the Planck length $l_P = G^{1/2}$, where $G$ is the gravitational constant. However, there is no general agreement on the detailed form of these inequalities and on the operational definition of the quantities $\Delta x^{\alpha}$.

One of the numerous approaches to this problem \cite{Ferretti, NVD, DFR, Camelia} is based on the statement that small values of $\Delta x^{\alpha}$ can be obtained only if there is a high concentration of energy which generates a strong gravitational field and possibly singularities of the metric, which are considered incompatible with the measurement procedure. In order to clarify these arguments, it is important to understand which values of $\Delta x^{\alpha}$ are compatible with a situation in which the gravitational fields are weak and the non-linear features of general relativity are not relevant. In other words, one would like to know which precision can be attained in a measurement procedure which can be described with a good approximation by means of a Poincar\'e covariant quantum theory. Of course, the fact that a measurement procedure cannot be described by the known theories does not mean that the measurement is impossible and a more detailed analysis is necessary. In the following we indicate how the results obtained in the preceding Sections can help to clarify these problems.

We consider a two-particle system. In order to have small values of $\Delta x^{\alpha}$, the CoM energy $\mu$ must be very large and we can disregard the masses of the particles. In the CoM system the particles have a momentum $\mu/2$ and the impact parameter is $b = 2j/\mu$. If $b$ is small, the two particles come very close to each other and, since their energy is very large, they may have a strong gravitational interaction. This can be avoided if $b$ and the angular momentum $\hbar j$ are sufficiently large (we have temporarily reintroduced $\hbar$ in order to distinguish classical and quantum effects). If we disregard quantum effects and the masses of the particles, the angular momentum has to be compared with a function of the CoM energy $\mu$ and the gravitational constant $G$. The only function with the right dimension is $\mu^2 G$ and we see that the required condition has the form (we put again $\hbar = 1$)
\begin{equation} \label{Peripheral}
j \gg \mu^2 G.
\end{equation}
It would be istructive to replace this dimensional argument by a detailed analysis.

Then from eq.\ (\ref{Ineq}) we find that a measurement of the coordinates of a two-particle collision cannot be described by a Poincar\'e covariant quantum theory unless we have
\begin{equation}
\sum_{r=1}^3 (\Delta x^r)^2 \gg G = l_P^2.
\end{equation}
We have already remarked that this does not mean that a more precise measurement is impossible.

It also follows from the reasults of Section 3 that the condition  (\ref{Peripheral}) does not imply any restriction on the quantities $\Delta x^0$ and, for instance, $\Delta x^3$. In this case too we have to be careful, because it is not sure that the POVMs that permit arbitrarily small values of $\Delta x^0$ and $\Delta x^3$ correspond to physical measurement procedures. In other words, we could have disregarded some relevant physical condition. Moreover, the space-time uncertainty relations can also be introduced by means of arguments which are not based on the unwanted appearance of strong gravitational fields \cite{Mead, JR5}.

We shall discus these problems with more detail elsewhere. Here our aim is just to suggest that the inequality (\ref{Ineq}) may play a role in the discussion of the space-time uncertainty relations due to quantum gravity.


\newpage

\begin{thebibliography}{999} 

\bibitem{JR} M. T. Jaekel and S. Reynaud:
{\it Space-Time Localization with Quantum Fields.}
Phys. Lett. {\bf A 220} (1996) 10. 

\bibitem{JR2} M. T. Jaekel and S. Reynaud:
{\it Mass as a Relativistic Quantum Observable.}
Europhys. Lett. {\bf 38} (1997) 1.

\bibitem{JR3} M. T. Jaekel and R. Reynaud:
{\it Conformal Symmetry and Quantum Relativity.}   
Found. Phys. {\bf 28} (1998) 437.

\bibitem{JR4} M. T. Jaekel and S. Reynaud:
{\it Quantum Localisation Observables and Accelerated Frames}
Preprint quant-ph/9806097 (1998).   

\bibitem{Toller} M. Toller: 
{\it Localization of Events in Space-Time.}
Preprint quant-ph/9702060 (1997).

\bibitem{Wightman} A. S. Wightman:
{\it On the Localizability of Quantum Mechanical Systems.}
Rev. Mod. Phys. {\bf 34} (1962) 845. 

\bibitem{Pauli} W. Pauli:
{\it Die allgemeinen Prinzipien der Wellenmechanik.}
Handbuch der Physik, edited by S. Fl\"ugge, vol. V/1, p. 60, Springer Verlag, Berlin, 1958.  

\bibitem{Davies} E. B. Davies:
{\it Quantum Theory of Open Systems.}   
Academic Press, London, (1976).       

\bibitem{Holevo} A. S. Holevo:
{\it Probabilistic and Statistical Aspect of Quantum Theory.}
North Holland, Amsterdam (1982).       

\bibitem{Werner} R. Werner:
{\it Screen Observables in Relativistic and Nonrelativistic Quantum Mechanics.}
J. Math. Phys. {\bf 27} (1986) 793.

\bibitem{BLM} P. Busch, P. J. Lahti and P. Mittelstaedt:
{\it The Quantum Theory of Measurement.}
Lecture Notes in Physics m2, Springer Verlag, Berlin (1991).

\bibitem{BGL} P. Busch, M. Grabowski and P. J. Lahti:
{\it Time Observables in Quantum Theory.}
Phys. Lett. {\bf A 191} (1994) 357. 

\bibitem{BGL2} P. Busch, M. Grabowski and P. J. Lahti:
{\it Operational Quantum Physics.}
Springer Verlag, New York (1995).   
                                                  
\bibitem{Giannitrapani} R. Giannitrapani:
{\it Positive-Operator-Valued Time Observable in Quantum Mechanics.}
Int. Journ. Theor. Phys. {\bf 36} (1997) 1575.     

\bibitem{JW} M. Jacob and G. C. Wick:
{\it On the General Theory of Collisions for Particles with Spin.}
Ann. Phys. (N. Y.) {\bf 7} (1959) 404.

\bibitem{Wigner} E. P. Wigner:
{\it On Unitary Representations of the Inhomogeneous Lorentz Group.}
Ann. of Math. {\bf 40} (1939) 149.                                   

\bibitem{Naimark} M. A. Naimark:
{\it Linear Representations of the Lorentz Group.}
Pergamon Press, London (1964).

\bibitem{GGV} I. M. Gel'fand, M. I. Graev and N. Ya. Vilenkin:
{\it Generalized Functions, Vol. 5.}
Academic Press, New York (1966).

\bibitem{Ruhl} W. R\"uhl:
{\it The Lorentz Group and Harmonic Analysis.}
Benjamin, New York (1970).

\bibitem{Garay} L. J. Garay:
{\it Quantum Gravity and Minimum Length.}
Int. J. Mod. Phys. {\bf A 10} (1995) 145.

\bibitem{Ferretti} B. Ferretti:
{\it On the Existence of a Minorant of the Indefiniteness for the 
Measurement of a Position.}
Lett. Nuovo Cimento {\bf 40} (1984) 169. 

\bibitem{NVD} Y. J. Ng and H. Van Dam:
{\it Limit to Space-Time Measurement.}
Mod. Phys. Lett. {\bf A 9} (1994) 335.
  
\bibitem{DFR} S. Doplicher, K. Fredenhagen and J. E. Roberts:
{\it The Quantum Structure of Spacetime at the Planck Scale and Quantum 
Fields.}
Commun. Math. Phys. {\bf 172} (1995) 187.  

\bibitem{Camelia} G. Amelino-Camelia:
{\it On Local Observables in Quantum Gravity.}   
Mod. Phys. Lett. {\bf A 11} (1996) 1411.   

\bibitem{Mead} C. A. Mead:
{\it Possible Connection between Gravitation and Fundamental Length.}
Phys. Rev. {\bf 135 B} (1964) 849. 

\bibitem{JR5} M. T. Jaekel and S. Reynaud:
{\it Gravitational Quantum Limit for Length Measurement}
Phys. Lett. {\bf A 185} (1994) 143. 

\end{thebibliography}
\end{document}